\documentclass[aps,prl,twocolumn,floatfix,showpacs,a4paper]{revtex4-1}
\usepackage{graphicx}
\usepackage{amssymb}
\usepackage{amsmath}
\usepackage{bookmark}
\usepackage{hyperref}

\renewcommand{\phi}{\varphi}
\newcommand{\paramTransitionVoltage}{ 7 V }

\begin{document}

\title{Observation of Quadrupole Transitions and Edge Mode Topology in an LC network}
\author{Marc Serra-Garcia}
\author{Roman S\"usstrunk}
\author{Sebastian D. Huber}
\affiliation{Theoretische Physik, ETH Zurich, CH-8093
Zurich, Switzerland}
\date{\today}

\begin{abstract}
High-order topological insulators are a recent development extending the topological theory of charge polarization to higher multipole moments. Since their theoretical proposal, several experimental realizations of high-order topological insulators have been reported. However, high order topological transitions have not been observed. In this letter, we report on the observation of a high-order topological transition in a quadrupole topological insulator implemented in an LC circuit with nonlinear couplings. This system presents the ability to confine electromagnetic energy in its corner states, with a localization length that can be tuned over a broad range through the use of an external bias voltage. Additionally, we provide an experimental characterization and an effective theory for the boundary states, further corroborating their topological nature by direct measurement of the winding number. 
\end{abstract}

\maketitle

A remarkable number of topological phenomena in condensed matter systems \cite{RyuClassificationSymmetries, KlitzingQuantizedHall, KonigSeTeQW, HsiehDiracInsulator, XuWeylSemimetal} can be explained as a consequence of a non-zero Berry phase \cite{BerryPhase, VanderbiltWannierCenters, VanderbiltTheoryPolarization} accumulated around closed trajectories in momentum space. Under suitable symmetries, this Berry phase is quantized \cite{ThoulessQuantizedConductance,KaneZ2}, establishing a sharp distinction between phases with different topological indices. Microscopically, the Berry phase can be expressed in terms of a momentum-dependent dipole moment, and manifests itself in surface phenomena such as topological edge states. Extending the theory of topological insulators to account for phenomena arising from higher-order (quadrupole, octopole) multipolar moments has been a long-standing problem, which has recently been solved theoretically by Benalcazar and collaborators \cite{BernevigScience} through the recursive use of Wilson loop operators.

\begin{figure}[b]
\centering
\includegraphics{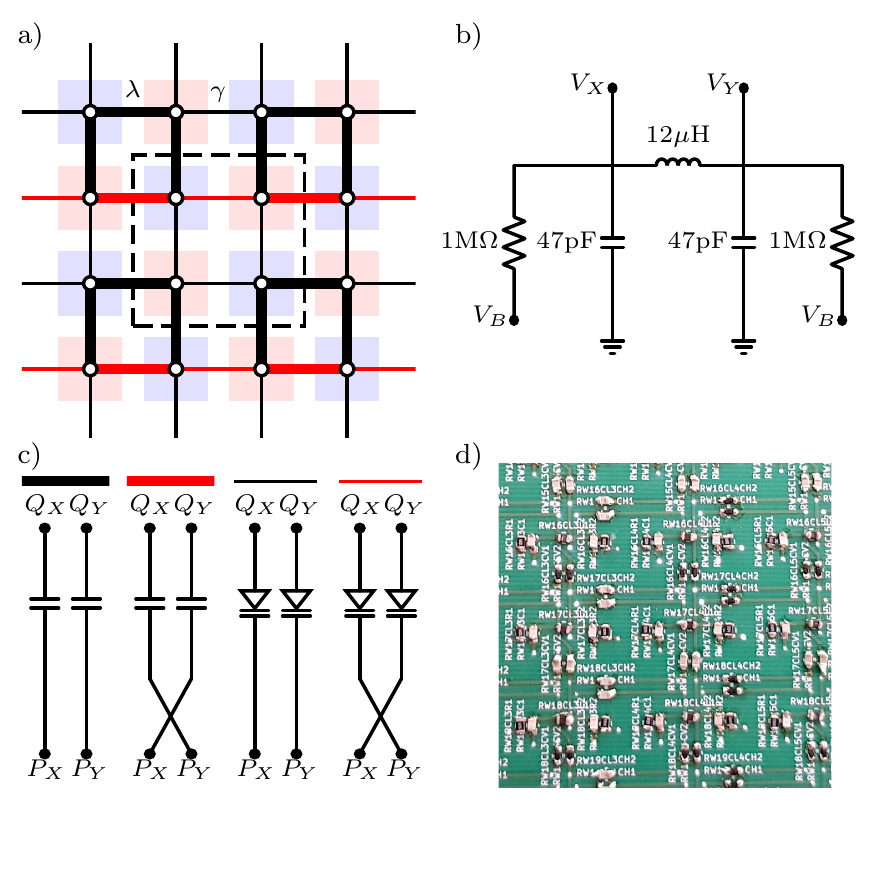}
\caption{Circuit model of the quadrupole topological insulator. a) Tight binding model of the material. Lines denote inter-site hoppings, with the thickness representing the hopping strength (in the topological phase), and the color indicating whether the hopping is positive (black) or negative (red). The background checkerboard pattern indicates the DC bias applied to each site, with blue denoting negative bias ($B_-$) and red denoting positive bias ($B_+$). b) Electrical schematic of the LC that constitutes each lattice site. c) Circuit implementations of the inter-site hoppings. d) Image of the printed circuit board displaying a full unit cell of quadrupole topological insulator. }
\label{fig:lion}
\end{figure}
A distinctive feature of topological insulators is the presence of boundary states arising from the topological properties of the bulk. In conventional topological insulators, a two-dimensional bulk gives rise to one-dimensional edge states, and a three-dimensional bulk gives rise to two-dimensional surface states. Quadrupole topological systems present boundary states that are themselves topologically nontrivial, nucleating corner modes in two-dimensional and edge modes in three-dimensional systems. This phenomenology is central to potential applications of high-order topological insulators as topologically protected waveguides in three dimensional systems. A remaining issue after the discovery of the high-order topological insulator is the identification of physical systems presenting high-order topological properties.

A straightforward route towards the experimental realization of high-order topological insulators consists in identifying a tight-binding model with suitable properties \cite{BernevigScience, BergholtzTightBindingModels} and translating it into a physical system through the use of discrete components \cite{SusstrunkPendulaScience, SimonSchusterCircuit, GlazmanCircuit} or a metamaterial geometry \cite{MatlackPerturb, SerraGarciaQuadrupole}. An example of such a tight-binding model for a quadrupole topological insulator is presented in Fig.~\ref{fig:lion}a \cite{BernevigScience}. This example combines non-commuting $M_x$ and $M_y$ mirror symmetries with a $\pi$ flux per plaquette, which provides the required ingredients for a quadrupole insulator, see Fig.~\ref{fig:lion}a. Shortly after their theoretical proposal, several experimental observations of higher-order topological effects have been reported in phononic \cite{SerraGarciaQuadrupole}, electronic \cite{SchindlerBismuth}, microwave \cite{BahlMicrowaveQuadrupole} and circuit \cite{Lee17, TomaleTopoelectricalQuadrupole} models. However, phase transitions in higher-order topological systems have not been experimentally observed or characterized.
  
In this paper we report an observation of a higher-order topological transition in a quadrupole topological insulator implemented in a LC circuit model, following conventional strategies to extend the concepts of charge polarization to photonic \cite{LuTopologicalPhonons} and phononic \cite{SusstrunkPendulaScience, SerraGarciaQuadrupole} metamaterials. We demonstrate the topological nature of the edge states in the non-trivial phase by direct measurements of the boundary winding number. 

LC circuit models are well suited to the implementation of tight binding models due to the fact that commercial capacitors and inductors closely resemble an ideal single degree-of-freedom system over a broad range of frequencies. Therefore, it is possible to establish a correspondence between individual hopping terms in the tight binding model and individual components in its circuit realization. This potential has already been exploited to implement conventional \cite{SimonSchusterCircuit,GlazmanCircuit} and and higher-order \cite{TomaleTopoelectricalQuadrupole} topological insulators. Capacitors in LC networks can be replaced by nonlinear varicap diodes, whose linearized capacitance can be tuned over a large range.

To realize the system experimentally, we map each degree of freedom in the tight-binding model (Fig.~\ref{fig:lion}a) into an LC resonator configured in a $\pi$ topology (Fig.~\ref{fig:lion}b), with an inductance $L=12\,\mu{\rm H}$ and a capacitance $C=47\,{\rm pF}$. This circuit presents two degrees of freedom: A zero frequency mode with the terminals $V_X$ and $V_Y$ oscillating in phase, and a mode with natural frequency $\omega_0=\sqrt{2/LC}$, where the two terminals oscillate with a phase difference of $\pi$. It makes sense to express the state of each resonant circuit in its eigenmode basis, with $\psi = V_X-V_Y$ and $\phi = V_X + V_Y$.  With these definitions, $\psi$ is taken as the dynamical degree of freedom, while the $\phi$ is used as an additional channel carrying the DC bias voltage responsible for tuning the inter-site hopping potential. The bias voltage $\phi$ is applied through two resistors connected to positive ($V_{B+}$) or negative ($V_{B-}$) bias planes alternating in a checkerboard pattern (Fig.~\ref{fig:lion}a). 

The coupling between unit cells is accomplished by means of symmetric capacitive interconnections, which can be twisted to achieve negative couplings (Fig.~\ref{fig:lion}c). Fixed couplings (represented by thick lines in Fig.~\ref{fig:lion}a) are implemented using $\lambda = 18\,{\rm pF}$ capacitors, while the variable interconnects (Thin lines in Fig.~\ref{fig:lion}a) are implemented using non-linear varicap diodes (BB659H7902) whose linearized capacitance can be swept between $\gamma = 2.5\,{\rm pF}$ and $\gamma = 38\,{\rm pF}$, decreasing nonlinearly with increased bias voltage. The system will be in the topological phase when $\lambda > \gamma$. 

Our experimental sample consists of a 18x8 sites lattice, containing 9x4 unit cells. The circuits is built on a 334x259 mm, four-layer printed circuit board (PCB), where the bottom layer is the ground, the top layer contains the signal routing, and the two middle layers are the $V_{B+}$ and $V_{B-}$ bias planes. The circuit is measured using a custom-made scanning probe with an input capacitance below $0.5\,{\rm pF}$ to prevent disturbing the circuit during measurements. 

The lattice is driven from a signal generator through a $1\,{\rm pF}$ capacitor, applied either at the measurement point (resulting in a measured quantity proportional to $\psi^2$ or at a different site, providing a phase sensitive Green's function measurement. 

\begin{figure}[t]
\centering
\includegraphics{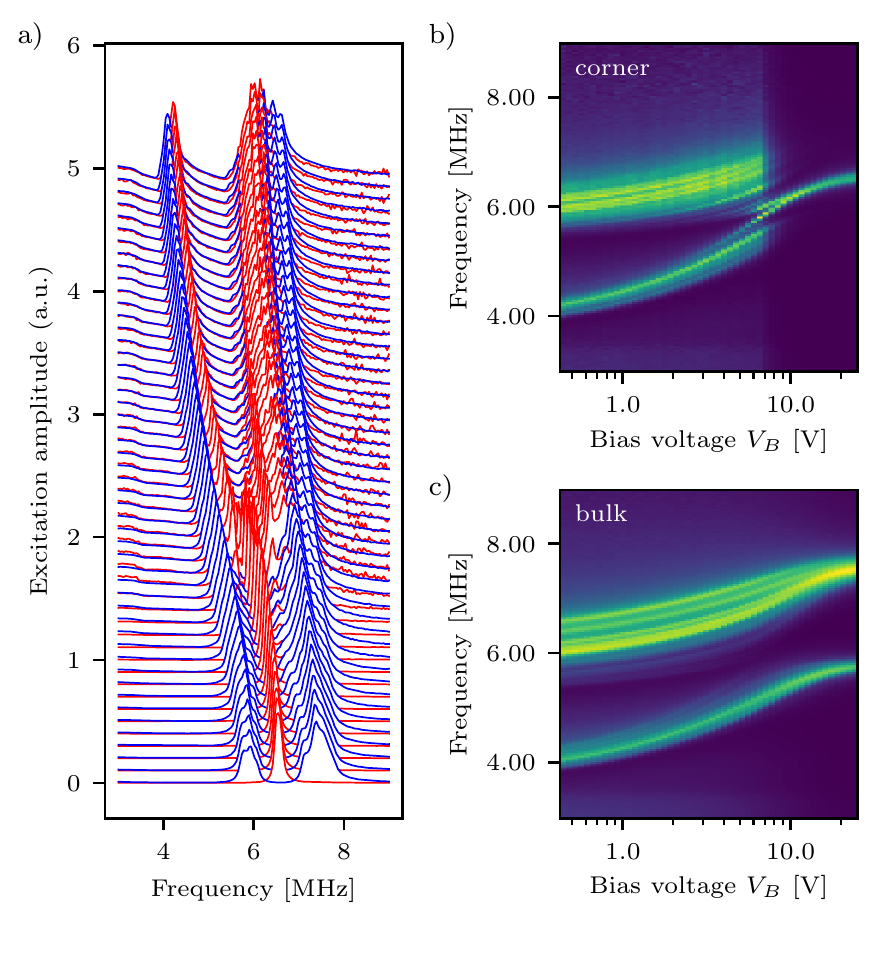}
\caption{Nucleation of corner modes and topological transition. a) Mean response $\langle \psi^2\rangle$ of the lattice in the bulk (blue) and corners (red). b) Corner response $\psi^2_{Corner}$ as a function of the excitation frequency and bias voltage $V_{B+}-V_{B-}$. c) Bulk response $\psi^2_{Bulk}$ as a function of the excitation frequency and bias voltage $V_{B+}-V_{B-}$. }
\label{fig:pt}
\end{figure}

Figure~\ref{fig:pt} depicts the distribution of energy as the bias voltage is ramped between $0\,{\rm V}$ and $24\,{\rm V}$. For voltages below \paramTransitionVoltage, $\lambda < \gamma$ and the system is in a trivial phase. In this phase, the system presents a gapped response with no edge or corner-dominated frequency region. When the bias voltage reaches \paramTransitionVoltage, we observe a bulk gap closing (Fig.~\ref{fig:pt}c) and the system enters the topological phase. 

Consistently with the theory of quadrupole topological insulators, this phase is characterized by mid-gap states that are localized at the corner (Fig.~\ref{fig:pt}b). For biases close to the transition voltage, we observe a non-negligible splitting in the corner response (Fig.~\ref{fig:pt}b). This splitting occurs due to hybridization between corner states: As a consequence of the finite size of the sample and the corner mode's localization length, there is non-zero overlap between modes at opposite corners, resulting in a non-zero corner--corner hybridization.

\begin{figure}[t]
\centering
\includegraphics{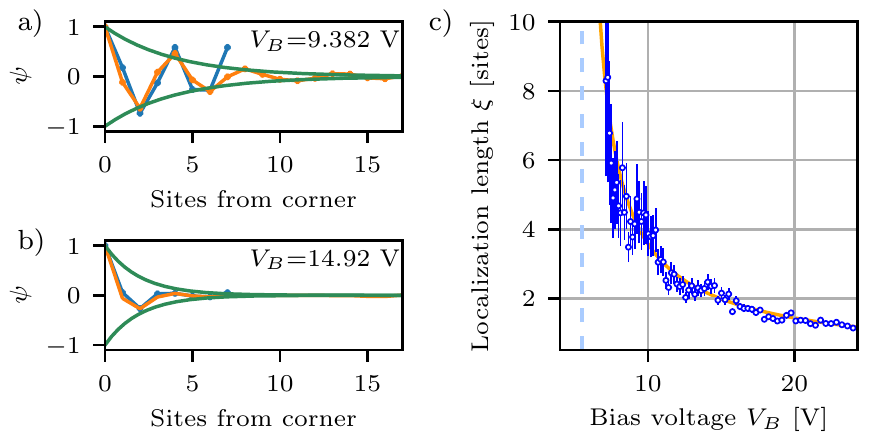}
\caption{Tuning the corner mode localization length. a) and b) show the measured normalized corner mode Green's function at low a) and high b) bias voltages. The orange and blue lines are measured along a long and short edges of the sample, respectively. The blue line corresponds to the prediction of the effective Su-Schrieffer-Heeger model. c) Measured localization length $\xi$ as a function of the bias voltage (dots) and theoretical prediction (orange line). The dashed line corresponds to the theoretically predicated phase transition point.}
\label{fig:decay}
\end{figure}

The boundary physics of a quadrupole topological insulator can be understood as an emergent Su-Schrieffer-Heeger (SSH) model \cite{SSH}.
We can obtain an effective description for the surface physics by considering states localized at the edge of a semi-infinite system ($-\infty < i < \infty$ and $0 \leq j < \infty$). These states are of the form $\psi(i,j)=\psi(i)\cos({\pi j}/2)e^{-j/{\xi}}$ with $\xi = 2/\log(\lambda/\gamma)$ and are orthogonal to the bulk states. When expressed in terms of the states $\psi(i,j)$ and after taking a Fourier transform along the edge, the emergent SSH model reads
\begin{equation}
        \label{eqn:ssh}
        H=\sum_k (a_k^*, b_k^*) \begin{pmatrix} 
        -\omega_0^2 & \gamma+\lambda e^{ik} \\
        \gamma+\lambda e^{-ik} & -\omega_0^2 \end{pmatrix}
        \begin{pmatrix}
            a_k \\ b_k
        \end{pmatrix},
\end{equation}
where we $a_k$ and $b_k$ denote the weights on the different sub-lattices and $k$ is measured in inverse lattice constants. Note, that owing to the perfect decoupling provided by $\psi(i,j)$ the emergent edge theory has the same hopping strengths $\lambda$ and $\gamma$ that compose the bulk quadrupole system. 

This edge system presents its own localized boundary states when the edge presents a discontinuity, and these correspond to the corner modes of the quadrupole system. However, the two edges that intersect in a corner do not give rise to two localized mode, but a single one, as expected from the quadrupole theory.

The corner modes in the emerging boundary SSH also have a localization length given by $\xi = 2/\log(\lambda/\gamma)$. Therefore, our ability to tune  the $\gamma$ coupling by modifying the bias voltage translates into an ability to tune the localization length of the boundary modes \cite{LydonTunableLocalization} (Fig.~\ref{fig:decay}). Close to the phase transition, corner modes extend significantly along the sample's edges (Fig.~\ref{fig:decay}a), while at larger bias voltages (Fig.~\ref{fig:decay}b) the modes are highly localized. The experimentally measured localization length, which can be varied by a factor of 10, is in good agreement with the theoretical prediction (Fig. ~\ref{fig:decay}c). The localization results corroborate the proposed explanation on the corner mode splitting observed in Fig.~\ref{fig:pt}b: The bias voltage below which the splitting occurs ($V_B \approx 10\,{\rm V}$), corresponds to the point where the corner mode localization length grows above 4 unit cells, the smallest dimension of our sample. These results are remarkable because they provide a mechanism to switch and control interactions between topologically-protected modes, for example for information-processing applications.

\begin{figure}[t]
\centering
\includegraphics{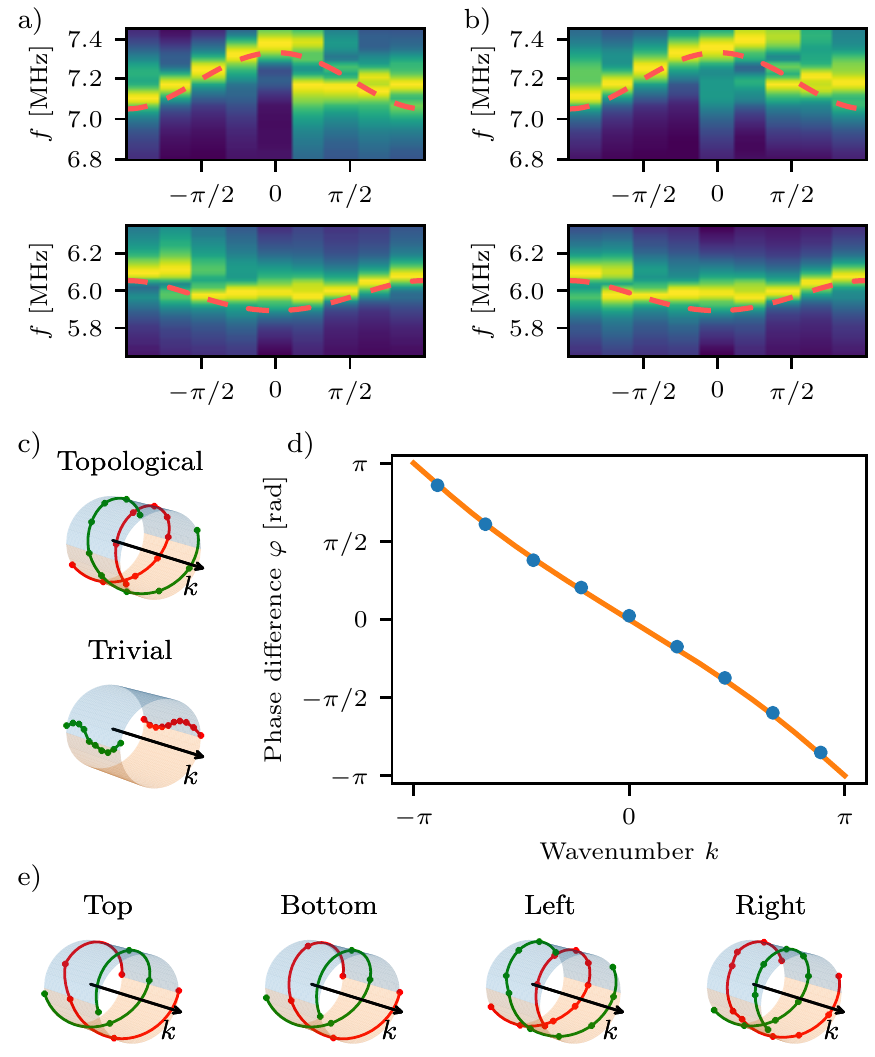}
\caption{Characterization of the edge mode topology. a) and b) Momentum-space response $\psi(k)$ of the boundary sites along a long edge of the lattice, calculated at odd a) and even b) sites. The orange dashed line corresponds to the theoretical prediction from the effective boundary theory. c) Winding diagram with the simulated phase response $\arg[\psi_{a}(k)/\psi_{b}(k)]$ of a topological (top) and trivial (bottom) emergent Su-Schrieffer-Heeger system, on the edge of a bulk quadrupole lattice, calculated above (red) below (green) the bandgap. d) Measured phase response $\arg[\psi_{a}(k)/\psi_{b}(k)]$ along a long edge of the material. e) Winding diagrams with the measured phase responses $\arg[\psi_{a}(k)/\psi_{b}(k)]$ along the four edges of the quadrupole topological insulator.}
\label{fig:topo}
\end{figure}

Emergent surface modes in topological systems have been studied theoretically, for example through Wilson loops \cite{FidkowskiWilsonLoops} or transfer function techniques based on recursively adding boundary elements to a finite system \cite{VonOppenEdgemodes}. Since our experimental system allows us to directly measure the boundary Green's function, we can experimentally characterize the topology of the edge states. 

In a quadrupole topological insulator, the emergent boundary system is of the SSH \cite{VonOppenEdgemodes} type. As for each $k$, Eq.~(\ref{eqn:ssh}) corresponds to a pseudo-spin in a ``magnetic field'' in the $xy$-plane, the eigenfunctions are of the form $\Phi(k)=(1, e^{i\varphi(k)})$. Since eigenfunctions must be identical at both edges of the Brillouin zone, $\varphi(k)$ can only vary in multiples of $2\pi$, with the number of $2\pi$ windings being the characteristic topological invariant.

To estimate the eigenfunctions $\Phi(k)$ and determine the boundary winding number, we require the boundary Green's function $\psi(k,\omega)$ as a function of the wavenumber $k$.  This is obtained by calculating the spatial Fourier transform of the boundary response. We consider even and odd sub-lattices separately. The resulting functions $a(k,\omega)$ and $b(k,\omega)$ present peaks that are in good agreement with the theoretical predictions for the boundary of SSH (Fig.~\ref{fig:topo}a,b). The winding number is determined by plotting the relative phase between even and odd sub-lattices, i.e, $\arg(a_k/b_k)$, (Fig.~\ref{fig:decay}c-e) at resonance, which corresponds to $\varphi(k)$, and then manually determining how many times it winds when crossing the Brillouin zone. 

We validate our method using numerical simulations of topological and trivial SSH systems (Fig.~\ref{fig:topo}c) and we confirm that it is extremely robust against damping. We then compare the experimental measurements of $\varphi(k)$ obtained along a long edge of our experimental systems to the theoretical value for the boundary SSH system (Fig.~\ref{fig:topo}d), and we observe a very good agreement with a winding of $2\pi$, indicating the topological nature of the phase. All boundary bands present nontrivial winding (Fig.~\ref{fig:topo}e), with the hopping sign prescribing the winding direction. The non-zero winding number along edge bands demonstrates that the edge system is itself topological, as expected from the theory of quadrupole systems \cite{BernevigScience,VonOppenEdgemodes}, and also corroborates the topologically-protected nature of the in-gap corner states. It should be noted that it is not possible to provide an equivalent discussion for the edge of the trivial phase since its response is not distinguishable from the bulk \cite{RosenthalCircuitWinding}.

We have reported on the observation of a topological transition in a quadrupole LC circuit model. In the topological phase, our system presents all the landmarks of quadrupole topological systems, including gapped edge modes and mid-gap corner states. The localization length of the corner states can be dynamically tuned by applying an external bias, enabling devices with localized, topologically protected modes with a variable profile. Our work also includes a direct measurement of the boundary topological invariants, which is remarkably robust against large amounts of damping. This work is the first investigation of a high-order topological system with nonlinear elements. While we have focused on the regime of small oscillations around a large constant voltage, future investigations ought to clarify the nonlinear effects that arise in the high oscillation amplitude regime.

\bibliographystyle{phd-url}
\bibliography{ref}

\end{document}